\documentclass[11pt]{article}
\usepackage{graphics,psfrag}
\usepackage{amssymb,epsfig,amsmath,euscript,array,cite}

\newcounter{multieqs}

\newcommand{\be}{\begin{equation}}
\newcommand{\ee}{\end{equation}}
\newcommand{\eq}[1]{(\ref{#1})}

\newcommand{\bra}[1]{\langle #1|}
\newcommand{\ket}[1]{|#1 \rangle}

\newcommand{\bm}[1]{\mbox{\boldmath $#1$}}

\def\bd{\begin{document}}
\def\ed{\end{document}}
\def\nn{\nonumber}
\def\bea{\begin{eqnarray}}
\def\eea{\end{eqnarray}}
\let\bm=\bibitem
\let\la=\label

\def\npb#1#2#3{Nucl. Phys. {\bf{B#1}} #3 (#2)}
\def\plb#1#2#3{Phys. Lett. {\bf{#1B}} #3 (#2)}
\def\prl#1#2#3{Phys. Rev. Lett. {\bf{#1}} #3 (#2)}
\def\prd#1#2#3{Phys. Rev. {D \bf{#1}} #3 (#2)}
\def\cmp#1#2#3{Comm. Math. Phys. {\bf{#1}} #3 (#2)}
\def\cqg#1#2#3{Class. Quantum Grav. {\bf{#1}} #3 (#2)}
\def\nppsa#1#2#3{Nucl. Phys. B (Proc. Suppl.) {\bf{#1A}}#3 (#2)}
\def\ap#1#2#3{Ann. of Phys. {\bf{#1}} #3 (#2)}
\def\ijmp#1#2#3{Int. J. Mod. Phys. {\bf{A#1}} #3 (#2)}
\def\rmp#1#2#3{Rev. Mod. Phys. {\bf{#1}} #3 (#2)}
\def\mpla#1#2#3{Mod. Phys. Lett. {\bf A#1} #3 (#2)}
\def\jhep#1#2#3{J. High Energy Phys. {\bf #1} #3 (#2)}
\def\atmp#1#2#3{Adv. Theor. Math. Phys. {\bf #1} #3 (#2)}

\newcommand{\EQ}[1]{\begin{equation} #1 \end{equation}}
\newcommand{\AL}[1]{\begin{subequations}\begin{align} #1 \end{align}
\end{subequations}}
\newcommand{\SP}[1]{\begin{equation}\begin{split} #1 \end{split}\end{equation}}
\newcommand{\ALAT}[2]{\begin{subequations}\begin{alignat}{#1} #2 
\end{alignat}\end{subequations}}
\def\beqa{\begin{eqnarray}} 
\def\eeqa{\end{eqnarray}} 
\def\beq{\begin{equation}} 
\def\eeq{\end{equation}} 

\def\N{{\cal N}}
\def\sst{\scriptscriptstyle}
\def\thetabar{\bar\theta}
\def\Tr{{\rm Tr}}
\def\one{\mbox{1 \kern-.59em {\rm l}}}

\def\a{\alpha}      \def\da{{\dot\alpha}}  
\def\b{\beta}       \def\db{{\dot\beta}}  
\def\c{\gamma}  \def\C{\Gamma}  \def\cdt{\dot\gamma}  
\def\d{\delta}  \def\D{\Delta}  \def\ddt{\dot\delta}  
\def\e{\epsilon}        \def\vare{\varepsilon}  
\def\f{\phi}    \def\F{\Phi}    \def\vvf{\f}  
\def\h{\eta}  
\def\k{\kappa}  
\def\l{\lambda} \def\L{\Lambda}  
\def\m{\mu} \def\n{\nu}  
\def\o{\omega}  
\def\p{\pi} \def\P{\Pi}  
\def\r{\rho}  
\def\s{\sigma}  \def\S{\Sigma}  
\def\t{\tau}  
\def\th{\theta} \def\Th{\Theta} \def\vth{\vartheta}  
\def\X{\Xeta}  
\def\z{\zeta}  
  
\def\cA{{\cal A}} \def\cB{{\cal B}} \def\cC{{\cal C}}  
\def\cD{{\cal D}} \def\cE{{\cal E}} \def\cF{{\cal F}}  
\def\cG{{\cal G}} \def\cH{{\cal H}} \def\cI{{\cal I}}  
\def\cJ{{\cal J}} \def\cK{{\cal K}} \def\cL{{\cal L}}  
\def\cM{{\cal M}} \def\cN{{\cal N}} \def\cO{{\cal O}}  
\def\cP{{\cal P}} \def\cQ{{\cal Q}} \def\cR{{\cal R}}  
\def\cS{{\cal S}} \def\cT{{\cal T}} \def\cU{{\cal U}}  
\def\cV{{\cal V}} \def\cW{{\cal W}} \def\cX{{\cal X}}  
\def\cY{{\cal Y}} \def\cZ{{\cal Z}}

\def\ua{\underline{\alpha}}  
\def\ub{\underline{\phantom{\alpha}}\!\!\!\beta}  
\def\uc{\underline{\phantom{\alpha}}\!\!\!\gamma}  
\def\um{\underline{\mu}}  
\def\ud{\underline\delta}  
\def\ue{\underline\epsilon}  
\def\una{\underline a}\def\unA{\underline A}  
\def\unb{\underline b}\def\unB{\underline B}  
\def\unc{\underline c}\def\unC{\underline C}  
\def\und{\underline d}\def\unD{\underline D}  
\def\une{\underline e}\def\unE{\underline E}  
\def\unf{\underline{\phantom{e}}\!\!\!\! f}\def\unF{\underline F}  
\def\unm{\underline m}\def\unM{\underline M}  
\def\unn{\underline n}\def\unN{\underline N}  
\def\unp{\underline{\phantom{a}}\!\!\! p}\def\unP{\underline P}  
\def\unq{\underline{\phantom{a}}\!\!\! q}  
\def\unQ{\underline{\phantom{A}}\!\!\!\! Q}  
\def\unH{\underline{H}}  
  
\def\As {{A \hspace{-6.4pt} \slash}\;}  
\def\bs {{b \hspace{-6.4pt} \slash}\;}  
\def\Ds {{D \hspace{-6.4pt} \slash}\;}  
\def\ds {{\del \hspace{-6.4pt} \slash}\;}  
\def\ss {{\s \hspace{-6.4pt} \slash}\;}  
\def\ks {{ k \hspace{-6.4pt} \slash}\;}  
\def\ps {{p \hspace{-6.4pt} \slash}\;}  
\def\pas {{{p_1} \hspace{-6.4pt} \slash}\;}  
\def\pbs {{{p_2} \hspace{-6.4pt} \slash}\;}  
  
\def\Fh{\hat{F}}  
\def\Vh{\hat{V}}  
\def\Xh{\hat{X}}  
\def\ah{\hat{a}}  
\def\xh{\hat{x}}  
\def\yh{\hat{y}}  
\def\ph{\hat{p}}  
\def\xih{\hat{\xi}}  
  
\def\psit{\tilde{\psi}}  
\def\Psit{\tilde{\Psi}}  
\def\tht{\tilde{\th}}  
   
\def\At{\tilde{A}}  
\def\Qt{\tilde{Q}}  
\def\Rt{\tilde{R}}  
\def\Nt{\tilde{N}}  
  
\def\at{\tilde{a}}  
\def\st{\tilde{s}}  
\def\ft{\tilde{f}}  
\def\pt{\tilde{p}}  
\def\qt{\tilde{q}}  
\def\vt{\tilde{v}}  
\def\nt{\tilde{n}}  
  
\def\delb{\bar{\partial}}  
\def\bz{\bar{z}}  
\def\bD{\bar{D}}  
\def\bB{\bar{B}}  
  
\def\ba{{\bf a}} 
\def\bk{{\bf k}}  
\def\bl{{\bf l}}  
\def\bp{{\bf p}}  
\def\bq{{\bf q}}  
\def\br{{\bf r}}
\def\bt{{\bf t}}
\def\bu{{\bf u}}
\def\bv{{\bf v}}
\def\bx{{\bf x}}  
\def\by{{\bf y}}  
\def\bR{{\bf R}}  
\def\bV{{\bf V}}  

\def\va{{\vec a}}
\def\vp{{\vec p}}
\def\vq{{\vec q}}
\def\vx{{\vec x}}
\def\vu{{\vec u}}
\def\vv{{\vec v}}
\newcommand{\ov}[1]{\overrightarrow{#1}}
  
\def\d{\delta}\def\D{\Delta}\def\ddt{\dot\delta}  
  
\def\pa{\partial} \def\del{\partial}  
\def\xx{\times}  
\def\uno{\mbox{1 \kern-.59em {\rm l}}}    
  
\def\trp{^{\top}}  
\def\inv{^{-1}}  
\def\dag{{^{\dagger}}}  
\def\pr{^{\prime}}  
  
\def\rar{\rightarrow}  
\def\lar{\leftarrow}  
\def\lrar{\leftrightarrow}  
  
\newcommand{\0}{\,\!}      
\def\one{1\!\!1\,\,}  
\def\im{\imath}  
\def\jm{\jmath}  
  
\newcommand{\tr}{\mbox{tr}}  
\newcommand{\slsh}[1]{/ \!\!\!\! #1}  
  
\def\vac{|0\rangle}  
\def\lvac{\langle 0|}  
  
\def\hlf{\frac{1}{2}}  
\def\ove#1{\frac{1}{#1}}  

\def\Box{\square}  
\def\ZZ{\mathbb{Z}}  
\def\CC#1{({\bf #1})}  
\def\bcomment#1{}  
\def\bfhat#1{{\bf \hat{#1}}}  
\def\VEV#1{\left\langle #1\right\rangle}  

\newcommand{\ex}[1]{{\rm e}^{#1}} \def\ii{{\rm i}}

\newcommand{\lrbrk}[1]{\left(#1\right)}
\newcommand{\sfrac}[2]{{\textstyle\frac{#1}{#2}}}

\font\mybb=msbm10 at 12pt
\def\bb#1{\hbox{\mybb#1}}

\font\myBB=msbm10 at 18pt
\def\BB#1{\hbox{\myBB#1}}

\title{{\bfseries Locality, Causality and \\
Noncommutative Geometry}}

\author{{\bfseries Chong-Sun Chu} \\
Centre for Particle Theory and Department of Mathematics,\\
University of Durham, 
Durham, DH1 3LE, United Kingdom. \\
Email: Chong-Sun.Chu@durham.ac.uk 
\and
{\bfseries Ko Furuta} \\
Theoretical Physics Laboratory,
The Institute of Physical and Chemical \\Research 
(RIKEN),
2-1 Hirosawa, Wako, Saitama 351-0198, Japan.
\\ Email: furuta@riken.jp
\and
{\bfseries Takeo Inami}\\
Department of Physics, Chuo University, Kasuga, \\Bunkyo-ku, 
Tokyo, 112-8551, Japan.\\
Email: inami@phys.chuo-u.ac.jp }

\begin{document}
\maketitle
\begin{abstract}
{We analyse the causality condition in noncommutative field
  theory and show that the nonlocality of noncommutative interaction
  leads to a modification of the light cone to the light wedge. This
  effect is generic for noncommutative geometry.  
We also check that the 
usual form  of energy condition is violated and propose that a new form 
is needed in noncommutative spacetime.
On reduction from  light cone to light wedge, it looks like
the noncommutative dimensions are effectively washed out 
and suggests a reformulation of noncommutative field theory in terms of
lower dimensional degree of freedom.
This reduction of dimensions due to noncommutative
  geometry could play a key role in explaining  the holographic property of
  quantum gravity.
}
\end{abstract}

\section{Introduction}

The study of field theories in noncommutative space has attracted a
great deal of attention (see for example, the review
\cite{review1,review2}) after it was revealed that low-energy
description of string theory in the presence of a constant NS-NS
$B$-field background gives rise to a certain class of  field theory on
noncommutative space \cite{SW} with commutation relations $[x^\m,x^\n] =
i \th^{\m\n}$. One of the important problems in the study of
noncommutative field theories is to understand how to 
define observable quantities,
see for example \cite{EM0,EM1} for discussions about the construction of
the energy-momentum tensor, and \cite{obs1,obs2} for the construction of
gauge invariant observable in noncommutative gauge theory. In
noncommutative space, locality is broken and it is not clear whether one
can define local physical quantities in general.

In commutative field theories, microcausality enables us to define
local observables. Also, microcausality is one  of the basic
assumptions leading  to the notion of S-matrix.  The study of
microcausality will help us understand better the notion of locality 
in noncommutative  field theory.
In noncommutative space, translational invariance
is intact.  However Lorentz symmetry is broken down to  a
smaller sub-group. This fact leads the authors of \cite{AG} to   
argue that the notion of a light cone is generally modified to  
that of a light wedge.  A modified  microcausality
condition was proposed which state that 
\begin{equation} 
[\Phi(x), \Phi(0)]_\pm = 0 , \mbox{if $(x^0)^2 - (\vx_c)^2 <0$}, 
\end{equation} 
where $+(-)$ is for fermionic (or bosonic)
fields and $\vx_c$ includes only the commutative space direction.  
The relation between violation of causality and unitarity was studied in
\cite{AG2}.
Another version where the commutator is replaced by a star-commutator
has also been proposed in \cite{chaichian1}. See also \cite{LiaoS} where
microcausality condition for composite operators has been studied.
Several properties of noncommutative field theories such as CPT theorem
\cite{AG,chaichian1},
spectral relations and dispersion relations of the S-matrix
\cite{LiaoS,chaichian2} have been studied with a modified microcausality condition
adapted for a light wedge. Attempts to study noncommutative field theories 
from an axiomatic approach have also been made 
\cite{AG,chaichian1,franco}.

Although the argument based on the spacetime symmetry generally allows
the causal region to be enlarged to the light wedge, it is still
necessary to study the causal structure in concrete models and see if
and how the light cone is modified. 
For example, we may consider a commutative theory in $d=6$
\begin{equation}
\label{lor} S= \int d^6 x \left( \frac{1}{2}(\partial \phi)^2
-\frac{1}{2} m^2 \phi^2 + 
g  c^{\mu \nu} \phi \partial_\mu \phi \partial_\nu \phi \right).  
\end{equation} 
If we consider the case in which only  $c^{45} \neq 0$, 
then the Lorentz symmetry is
broken  to $SO(3,1)\times SO(2)$. One may expect that the light 
wedge will arise too due to the general analysis of \cite{AG}. 
However as will be clear from our analysis below, microcausality 
condition is still given by  the usual light cone for the theory
\eq{lor}. The crucial difference between a noncommutative theory 
and a generic Lorentz symmetry broken theory lies in the {\it nonlocality} 
of noncommutative interaction. 
This important aspect was missing in the general argument of \cite{AG}.

In this paper, we study  the microcausality in a concrete model, the
noncommutative $\phi^3$ theory. In section 2, we  show that due to the 
phase factor characteristic of the nonplanar diagram,  the 
light cone is  modified to the
light wedge in perturbative computation. Also it will be clear
from our computation that this modification is generic and holds for
general noncommutative field theory. In section 3 we  study the 
energy conditions in  noncommutative field theories and show that 
their usual form is violated. 
This is consistent with our result in section 2.
Discussion and conclusions are made in section 4.

\section{Microcausality condition in perturbative theory}

\subsection{Commutative case}

For a vector $v^\m$, we use the notation $\bv$ to denote its spatial part.
We have $u \cdot v = u^0 v^0 -\bu \cdot \bv$.
We also use  the light cone coordinates. 
Define $v_\pm = (v^0 \pm v^5)/\sqrt{2}$ and denote the
orthogonal part by $\vv$. We have 
$u \cdot v = u_+ v_- + u_- v_+ - \vu\cdot \vv$.

We start with the commutative case to see how 
the light cone condition is derived in a canonical quantization 
formalism.  Consider the $\phi^3$ theory in six dimensions.
To study the causal structure, we consider the matrix element of the 
field commutator
\begin{equation} \label{mat-ele}
\cM := \langle \alpha | [ \phi_H (x_1), \phi_H (x_2) ] |\beta\rangle.
\end{equation}
Here $\phi_H$ is the operator in the Heisenberg picture
\begin{equation}
\phi_H (x)=U^\dag (t,t_0 )\phi (x) U(t,t_0),
\end{equation}
$\phi$ is the field  operator in the interaction picture 
\begin{equation}
\phi(t,\vx)=\int \frac{d^5 \bp}{(2\pi)^5}
\frac{1} { \sqrt{2\omega_p} } 
\left(
a_p\; e^{-ip\cdot x} +a_p^\dag\; e^{ip \cdot x}\right)
\bigg|_{x^0=t-t_0}, \quad 
\o_p := \sqrt{\bp^2+m^2}
\end{equation}
and  $U(t_1,t_2)$ is the time evolution operator 
\begin{equation}
U(t_1,t_2)=T \exp \left(ig\int^{t_2}_{t_1} dt V(t)\right),
\quad 
V(t) =  \int d^5 \vx \; \frac{1}{3!} :\phi^3:.
\end{equation}
We use a normal ordered interaction term here. However this is
inessential to our argument, the same result will be obtained without
the normal ordering. We will make more comments on this at the end
of this section. For simplicity, we consider the states
\begin{equation}
|\alpha\rangle=|\beta\rangle=U^\dag (t_1,t_0 )| 0\rangle ,
\end{equation}
where $| 0\rangle$ is the perturbative vacuum $a_\bp | 0\rangle=0$.
Due to translational invariance, $\cM$ is a
function of  $x^\m := x_1^\m -x_2^\m$.

We evaluate the matrix element \eq{mat-ele} in  perturbation theory. Up to
second order in $g$, we have 
\begin{equation}
\cM = \cM^{(0)} + \cM^{(1)} + \cM^{(2)} + \cdots , 
\end{equation} 
where
\begin{equation}
\cM^{(0)} = \bra{0} [\phi(x_1), \phi(x_2)] \ket{0} ,
\end{equation}
\begin{equation}
\cM^{(1)} = i g  \int_{t_1}^{t_2} ds \bra{0} \;
\Bigl[ \phi(x_1) V(s) \phi(x_2)  - \phi(x_1)\phi(x_2)  V(s)  + {\rm c.c.}
\Bigr]  \ket{0},
\end{equation}
\bea
\cM^{(2)} = -g^2   \int_{t_1}^{t_2} ds_1  \int_{t_1}^{s_1} ds_2 \;
\bra{0} && \Bigl[ 
\phi(x_1) V(s_1)V(s_2)\phi(x_2) + \phi(x_1) \phi(x_2) V(s_2)V(s_1) \nn\\
&& - \phi(x_1) V(s_1) \phi(x_2) V(s_2) -{\rm c.c.}
\Bigr]  \ket{0} .
\eea
For the $\phi^3$ interaction, it is easy to see that $\cM^{(1)} =0$.
It is easy to show that $\cM^{(0)} =0$ iff 
\begin{equation}
\d x^2 := (x_1-x_2)^2   =2 x^+x^- - \vx^2 <0.
\end{equation} 
For completeness we include an elementary calculation in the appendix.

For $\cM^{(2)}$,  a straightforward calculation yields
\begin{equation} \label{M2-v}
\cM^{(2)}=\frac{-g^2}{(2\pi)^{10}}\int 
\frac{d^5 \bq_2 d^5 \bq_3}{2\omega_{q_2}
2\omega_{q_3}(2\omega_{p_1})^2}
e^{-i p_1 x}f-\mbox{c.c.} \; , 
\end{equation}
where
\begin{equation}
\bp_1= \bq_2+\bq_3, \qquad
 p_1^0 = \o_{p_1} = \sqrt{(\bq_2+\bq_3)^2 +m^2},
\end{equation}
\begin{equation} \label{f-def}
f=\frac{t}{2i\tilde{\omega}}+\frac{1}{2\tilde{\omega}^2}
+\frac{1}{\tilde{\omega}\hat{\omega}}-\frac{1}{\tilde{\omega}
\omega_{p_1}}+\frac{1}{\tilde{\omega}\hat{\omega}}
e^{i\tilde{\omega}t}-\frac{1}{2\tilde{\omega}^2}e^{-i\tilde{\omega}t},
\end{equation}
\bea
 & \hat{\omega}:= \omega_{q_2}+\omega_{q_3}+\omega_{p_1},\\
& \tilde{\omega} := -\omega_{q_2}-\omega_{q_3}+\omega_{p_1}.
\eea
The contributions leading to \eq{M2-v} are given in figure 1, the planar 
diagram. 

\begin{figure}[ht] \label{bubble}
\psfrag{p1}{\LARGE $p_1$}
\psfrag{q2}{\LARGE $q_2$}
\psfrag{q3}{\LARGE $q_3$}
\begin{center}
{\scalebox{0.7}{\includegraphics{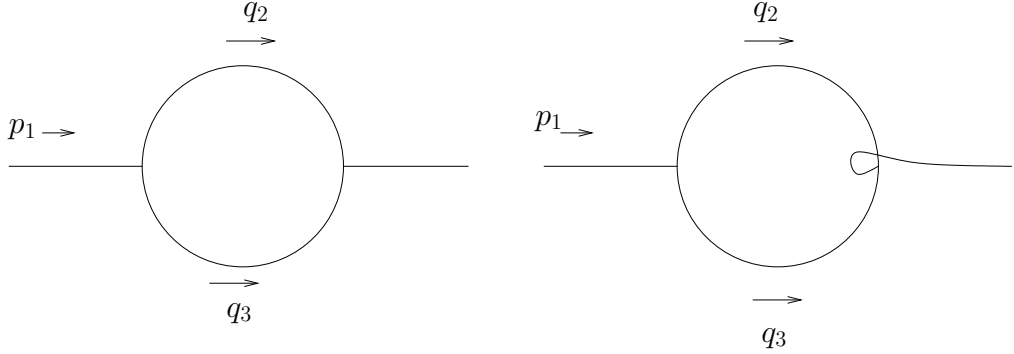}}}
\end{center}
\caption{Planar and nonplanar contributions }
\end{figure}

Using 
\begin{equation} \label{measure}
\int \frac{d^5 \bq}{\o_{q}} = 
\int_0^\infty \frac{d q^+}{q^+}  d^4 \vq 
\bigg|_{q^- = (\vq^2+m^2)/2 q^+},
\end{equation}
we obtain the light cone representation of $\cM^{(2)}$ as
\begin{equation} \label{M2-b}
\cM^{(2)}= g^2\left[
\int^\infty_0 d q^+_2 \int^\infty_0 d q^+_3 G(q_i^+,x)-
\int^\infty_0 d q^+_2 \int^\infty_0 d q^+_3 G(q_i^+,-x) \right],
\end{equation}
where
\begin{equation} \label{G}
G(q_i^+,x) := e^{-i\frac{\d x^2} {2x^+}(q_2^+ +q_3^+)
-i\frac{m^2 x^+}{2} \left( \frac{1}{q_2^+}+\frac{1}{q_3^+}\right) 
} H(q_i^+,x)
\end{equation}
and the kernel $H(q_i^+,x)$ is defined as
\begin{equation} \label{H}
H(q_i^+,x) :=
\int \frac{d^4 \vq_2 d^4 \vq_3 }{(2\pi)^{10}}\; 
e^{-i\frac{x^+}{2q_2^+}\left(\vq_2 -q_2^+\frac{\vx}{x^+}\right)^2}
e^{-i\frac{x^+}{2q_3^+}\left(\vq_3 -q_3^+\frac{\vx}{x^+}\right)^2}
\frac{f e^{-i\tilde{\omega}t}}{16 q_2^+ q_3^+ \o_{p_1}^2} .
\end{equation}
The frequencies $\o_{q_i}$ and $\o_{p_1}$ are to 
be written in terms of the light cone and transverse components. We have
\begin{equation} \label{freq-qi}
\o_{q_i} = \frac{1}{\sqrt{2}}(q_i^+ + \frac{1}{2 q_i^+} (\vq_i{}^2 +m^2)) ,
\quad i=2,3,
\end{equation}
\begin{equation} \label{freq-p1}
\o_{p_1}^2 =(\vq_2 + \vq_3)^2 + m^2+ 
\frac{1}{2}\biggl[q_2^+ + q_3^+ - \frac{1}{2q_2^+}(\vq_2{}^2+m^2) 
-\frac{1}{2q_3^+}(\vq_3{}^2+m^2)\biggr]^2 .
\end{equation}
In the above we have chosen to write $G$ in the form \eq{G} which 
will be convenient for our analysis. Note that the 
exponential part in  \eq{G} is kinematical and is independent 
of the interaction.
The details of dynamics of the theory 
enter through $H(q_i^+,x)$ from the last piece in \eq{H}.

The convergence properties of the $q_2^+, q_3^+$ integrals in \eq{M2-b} 
depend crucially on the sign of $\d x^2$. For large $q_i^+$, 
\begin{equation}
\o_{q_i} \sim q_i^+, \quad
\o_{p_1} \sim q_2^+ + q_3^+,\quad 
\tilde{\o} \sim 1/ q_i^+, \quad
\hat{\o} \sim q_2^+ + q_3^+ , 
\end{equation}
up to numerical proportional  constants. It follows that
$f$ and $H(q_i^+,x)$ grow at most
polynomially with $q_i^+$.
The large $q_i^+$ behaviour of $G(q_i^+,x)$ is thus dominated by
the exponential part, 
\begin{equation}
G(q_i^+,x) \sim e^{-i(q_2^+ +q_3^+)\frac{\d x^2}{2x^+}}, \quad
\mbox{for large $q_i^+$}.
\end{equation}
Without loss of generality
\footnote{
If $x^+<0$, one can consider $ [\phi(x_2), \phi(x_1)]$. Or if one want,
one can keep $x^+<0$ for the analysis. All one need to do is to
perform an opposite rotation of contours to the one in
\eq{contour-rot}:
\begin{equation} \label{contour-rot'} 
 q_i^+ \to -i q_i^+ \quad \mbox{in the first integral in \eq{M2-b}}, \quad
 q_i^+ \to i q_i^+ \quad \mbox{in the second integral in \eq{M2-b}},
\end{equation}
and the same analysis leads to the light cone condition $\d x^2 <0$.
}, 
we assume $x^+ >0$,
$G(q_i^+,x)$ thus vanishes exponentially at large $q_i^+$ on the upper 
half-plane if $\d x^2 < 0$.
It can also be verified that there is no pole enclosed
when the integration path is rotated.
Thus we can rotate the contours of the integration to the imaginary axes
as follows without getting any extra contribution:
\bea \label{contour-rot}
&& q_i^+ \to i q_i^+ \quad \mbox{in the first integral in \eq{M2-b}},\nn\\
&& q_i^+ \to -i q_i^+ \quad \mbox{in the second integral in \eq{M2-b}}.
\eea
And we obtain
\begin{equation}
\cM^{(2)}= g^2\left[
\int^\infty_0 d q^+_2 \int^\infty_0 d q^+_3 G(iq_i^+,x)-
\int^\infty_0 d q^+_2 \int^\infty_0 d q^+_3 G(-iq_i^+,-x) \right].
\end{equation}
Since the frequencies $\o_{q_i}, \o_{p_1}$ 
change sign under $q^+ \to -q^+$ \footnote{Although it does not follow 
immediately from \eq{freq-p1}, one should 
change the sign of all frequencies simultaneously under $q^+ \to -q^+$  
for analytically.}, $G(-iq_i^+,-x) =G(iq_i^+,x)$ and hence
the two terms in \eq{M2-b} are  identical:
\bea \label{comm}
 \cM^{(2)}=&& -\int^\infty_0 d q^+_2 \int^\infty_0 d q^+_3
e^{(q_2^+ +q_3^+)\frac{\d x^2}{2x^+} - \frac{m^2 x^+}{2}
\left( \frac{1}{q_2^+}+\frac{1}{q_3^+}\right)} H(i q_i^+,x)\nonumber\\
&+ &\mbox{ same term as above}. 
\eea
Therefore $\cM^{(2)}$ vanishes if the first term of (\ref{comm}) 
is finite, i.e.,
\begin{equation}
\int^\infty_0 d q^+_2 \int^\infty_0 d q^+_3
e^{(q_2^+ +q_3^+)\frac{ \d x^2}{2x^+}-\frac{m^2 x^+}{2}
\left( \frac{1}{q_2^+}+\frac{1}{q_3^+}\right)} H(i q_i^+,x) <\infty.
\label{q-int}
\end{equation}
As we showed above, $H(q_i^+,x)$ grows at most polynomially with
$q_i^+$, thus the integral converges at large $q_i^+$  
if   $\d x^2 <0$. For small $q_i^+$, we have the damping factor ($m^2
>0$ since the theory is not tachyonic) in the exponent, thus the
integral converges also at  $q_i^+ =0$.
Therefore we obtain the result that,  up to order  $g^2$, 
the light cone is not modified by interaction.

We remark that at the tree level, the matrix element takes a similar
form. See \eq{com-tree}. In fact the form \eq{M2-b} and \eq{G} is quite general and
universal. In general, the matrix element at a certain order of coupling 
takes the form 
\begin{equation} \label{M-gen}
\cM^{(n)} \approx g^n \left[ \int \prod_i dq_i^+ G(q_i^+,x) - 
\int \prod_i dq_i^+ G(q_i^+,-x) \right] , 
\end{equation}
where  
\begin{equation} \label{G-gen}
G(q_i^+,x) := \exp \Bigl(
-i\frac{\d x^2} {2x^+} \sum_{i\in I} q_i^+ 
-i\frac{m^2 x^+}{2} \sum_{i\in I} 1/q_i^+  
\Bigr)
H(q_i^+,x)
\end{equation}
and $H(q_i^+,x)$ is some kernel which contains the details 
of the interaction. 
The momentum integration ranges over all the independent momenta after imposing
momentum conservation at each vertex. 
The index $i$ in the sum ranges over a subset $I$ of them.
For example, for the diagram in
figure 2, the independent momenta to be integrated can be chosen to be
$\bq_2, \bq_3 \cdots, \bq_n$ and the set $I$ may be taken to be $\{2,3 \}$.
A similar argument can be applied  and one arrives at the same light cone.  
The generalization to theory with more couplings and masses is straightforward.
 
\begin{figure}[ht] \label{higher-order}
\psfrag{p1}{\LARGE $p_1$}
\psfrag{q2}{\LARGE $q_2$}
\psfrag{q3}{\LARGE $q_3$}
\psfrag{q4}{\LARGE $q_4$}
\psfrag{q5}{\LARGE $q_5$}
\psfrag{qn}{\LARGE $q_n$}
\psfrag{cdots}{\LARGE $\cdots$}
\begin{center}
{\scalebox{0.7}{\includegraphics{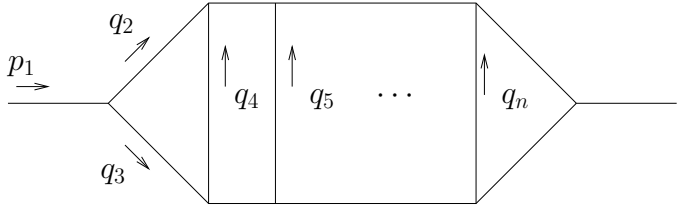}}}
\end{center}
\caption{Higher order contributions }
\end{figure}

\subsection{Noncommutative case}

We now turn to the noncommutative case. 
Since theory with noncommutativity in time 
has problem with unitarity \cite{u1}, see also \cite{AG2,u2},  
we will consider spatial noncommutativity
We can choose the light cone direction to be commutative.
\begin{align}
&\left[ x^i,x^j \right] =i\theta^{ij}, 
\quad i=1,2,3,4, \\
& \quad \theta^{\pm i}=\theta^{+ -}=0.
\end{align}
In noncommutative case, we can classify the terms contributing to $\cM^{(2)}$
into planar and nonplanar parts.
The planar parts give exactly the same terms (apart from numerical coefficient) 
as the commutative case.
For the nonplanar parts, we have the additional phase factor 
\begin{equation}
P(q_2,q_3)= e^{i \vq_2\theta \vq_3}.
\end{equation}

Consider the contribution coming from non-planar part. Up to a
numerical coefficient, 
\begin{equation}
\cM^{(2)}_{\rm np} \sim 
\int^\infty_0 d q^+_2 \int^\infty_0 d q^+_3
\left[ G_{\rm np}(q_i^+,x)-G_{\rm np}(q_i^+,-x) \right], \label{Mnp}
\end{equation}
where
\begin{equation} \label{Gnp}
G_{\rm np}(q_i^+,x) := e^{-i\frac{\d x^2} {2x^+}(q_2^+ +q_3^+)
-i\frac{m^2 x^+}{2} \left( \frac{1}{q_2^+}+\frac{1}{q_3^+}\right) 
} H_{\rm np}(q_i^+,x)
\end{equation}
and the kernel $H_{\rm np}$ is
\begin{equation} \label{Hnp}
H_{\rm np}(q_i^+,x) := 
\int \frac{d^4 \vq_2 d^4 \vq_3 }{(2\pi)^{10}}\; 
e^{-i\frac{x^+}{2q_2^+}\left(\vq_2 -q_2^+\frac{\vx}{x^+}\right)^2}
e^{-i\frac{x^+}{2q_3^+}\left(\vq_3 -q_3^+\frac{\vx}{x^+}\right)^2}  
\frac{f  e^{-i\tilde{\omega}t}}{16 q_2^+ q_3^+ \o_{p_1}^2}P(q_2,q_3),
\end{equation}
where $f$ is defined by \eq{f-def} above. 

To see the possibility of contour deformation,
we examine the large $q_i^+$ behaviour of $H_{\rm np}(q_i^+,x)$.
The $\vec{q}$-integral in $H_{\rm np}$ is of the form
\begin{equation}
H_{\rm np} (q_i^+,x) = \int \frac{d^4 \vq_2 d^4 \vq_3 }{(2\pi)^{10}} 
e^{A_2(\vq_2-\va_2)^2 +A_3(\vq_3-\va_3)^2 +i\vq_2\theta \vq_3}
 \frac{f e^{-i\tilde{\omega}t}}{ 16q_2^+q_3^+ \o_{p_1}^2} ,
\end{equation}
where $A_2=-\frac{i x^+}{2 q_2^+}$, $A_3=- \frac{i x^+}{2 q_3^+}$,
$\va_2=\frac{q_2^+ \vx}{x^+}$ and  $\va_3=\frac{q_3^+ \vx}{x^+}$.
As checked above,
$f e^{-i\tilde{\omega}t} /16q_2^+q_3^+\o_{p_1}^2$ diverges at large
$q_i^+$ at most polynomially. Thus we have for large $q_i^+$,
\begin{align}
H_{\rm np}&\sim\int d^4 \vq_2 d^4 \vq_3 e^{A_2(\vq_2-\va_2)^2
+A_3(\vq_3-\va_3)^2 +i\vq_2\theta \vq_3}\\
&= \int d^4\vq_2 e^{A_2(\vq_2-\va_2)^2}
\int d^4 \vq_3
\exp \biggl(
{A_3 \bigl( \vq_3+\frac{i\ov{q_2 \theta}-2A_3\va_3}
{2A_3}\bigr)^2+\frac{(\ov{q_2 \th})^2+4 i A_3\vq_2\theta
  \va_3}{4A_3}} \biggr) .
\label{J}
\end{align}
Now we choose $\theta^{ij}$ to be in the canonical form
\begin{equation}
\theta^{ij}=\begin{pmatrix} 
(\th')_{2\times 2} & \\
& (\th'')_{2\times 2}
\end{pmatrix} =
\begin{pmatrix} 
 & \theta' & & &\\ 
-\theta' & & & \\
 & & & \theta'' \\
 & & -\theta''& 
\end{pmatrix},
\end{equation}
then
\begin{equation}
(\ov{q_2 \th})^2=
\theta'{}^2  ({\vq_2}{}')^2 +\theta''{}^2 ({\vq_2}{}'')^2, 
\end{equation}
where
$\vec{q}_2{}'$ and $\vec{q}_2{}''$ represent the $1, 2$ 
components and $3, 4$ components of $\vec{q}_2$  respectively. 
The integral \eq{J} factorizes into a product of 2 dimensional integrals.
Integrating out $\vq_3$, we obtain 
\begin{align}
H_{\rm np}&\sim\int d^2\vq_2{}' \exp \left[
(\vq_2{}')^2  \left( A_2+\frac{{\theta'}^2} {4A_3}\right) 
-\vq_2{}' \cdot\left(2A_2\va_2 - i\ov{\theta'a_3}\right)
+A_2\va_2^2
\right]\nn \\
&\times\int d^2\vec{q}_2{}''
\exp\left[ \mbox{ $\theta'$ replaced by $\theta''$}\right].
\end{align}
This can be easily integrated and it gives
\begin{equation}
H_{\rm np} \sim e^{F'} e^{F''},
\end{equation}
where 
\begin{equation} 
F':= \frac{1}{E'}\left[ 
\frac{{\theta'}^2A_2} {4A_3}\vec{a}_2^2
+iA_2\vec{a}_2 \ov{\theta' a_3}+
\frac{(\ov{\theta' a_3})^2}{4}\right], 
\quad \mbox{and}\quad 
E':= A_2+\frac{{\theta'}^2}{4A_3}
\end{equation}
and a similar expression for $F''$ with $\th'$ replaced by $\th''$.
Substitute the expressions for $A_2, A_3, \va_2, \va_3$, we obtain
\begin{equation}
F' = -i \frac{\th'{}^2 \vx{}'{}^2 q_2^+ q_3^+(q_2^+ + q_3^+) }
{2 x^+(-x^+{}^2 + \th'{}^2 q_2^+ q_3^+)} 
\end{equation}
whose large $q_i^+$ behaviour can be read off easily
\begin{align}
&F'\to -\frac{i\vec{x}'{}^2}{2x^+}q_2^+ \mbox{ as } q_2^+\to \infty,\\
&F'\to -\frac{i\vec{x}'{}^2}{2x^+}q_3^+ \mbox{ as } q_3^+\to \infty.
\end{align}
Note that $\theta'$-dependence is cancelled completely!
Similarly, we have the same results for $F''$.
Thus for large $q_i^+$, the integral factor $H_{\rm np}(q_i^+,x)$ contributes
\begin{equation} \label{Hnc}
H_{\rm np}(q_i^+,x)\sim e^{-i\frac{\vec{x}_{nc}^2}{2x^+}(q_2^+ +q_3^+)}
\end{equation}
for each subspace of $\vec{x}_{nc}$ with non-zero noncommutativity. This
is in contrast to the commutative case where the dependence on $q_i^+$ is
subdominant compared to that of $G(q_i^+,x)$.
Because of \eq{Hnc},  the large $q_i^+$ behaviour of $G_{\rm
  np}(q_i^+,x)$ 
will be modified.  
Notice that the large $q_i^+$ behaviour of $H_{\rm np}$
matches the exponential factor of (\ref{Gnp}),
we have
\begin{align}
\cM^{(2)} &= {\rm coeff.} \times
\int^\infty_0 d q^+_2 \int^\infty_0 d q^+_3
\left[ e^{-i(q_2^+ +q_3^+)x^{-}-i\frac{m^2 x^+}{2}
\left( \frac{1}{q_2^+}+\frac{1}{q_3^+}\right) +i\frac{\vec{x}^2}
{2x^+}(q_2^+ +q_3^+)} H_{\rm np}(q_i^+,x)-\mbox{c.c.}
\right]\nonumber\\
&\sim\int^\infty_0 d q^+_2 \int^\infty_0 d q^+_3
\left[ e^{-i(q_2^+ +q_3^+)\frac{(\Delta x)^2}{2x^+}}-\mbox{c.c.}
\right],
\label{Mnp2}
\end{align}
where
\begin{equation}
(\Delta x)^2 := 2x^+ x^- -\vec{x}_c^2.
\end{equation}
$\vec{x}_c$ represents the commutative subspace of $\vec{x}$.
In the second line of (\ref{Mnp2}),
we have omitted factors which do not exponentially diverge
at large $q_i^+$.

Now we can repeat the same argument as in the commutative case:
\begin{align}
&\mbox{We can rotate the contour if }(\Delta x)^2<0; \nonumber\\
&\mbox{Integrals in (\ref{Mnp2}) converge and cancel out 
if }(\Delta x)^2<0.
\end{align}
Thus we see that the light cone ($(\delta x)^2<0$) is modified 
to light wedge ($(\Delta x)^2<0$) 
in order for the commutator to be zero.

\begin{figure}[ht] \label{uni}
\psfrag{a}{\huge (a)}
\psfrag{b}{\huge (b)}
\psfrag{c}{\huge (c)}
\psfrag{x1}{\huge  $x_1$}
\psfrag{x2}{\huge  $x_2$}
\begin{center} {\scalebox{0.55}{\includegraphics{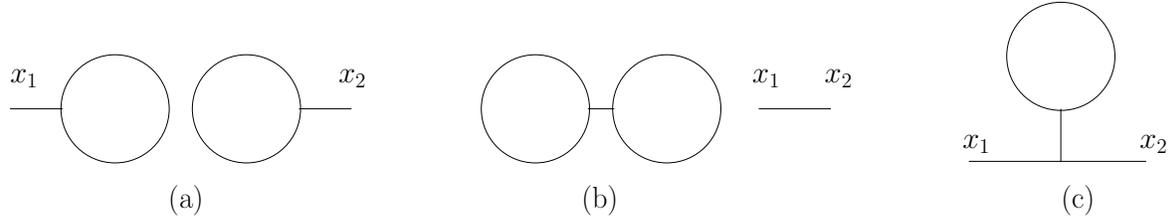}}} \end{center}
\caption{ Unimportant contributions }
\end{figure}

A couple of remarks are in order.

\begin{itemize} 
\item[{\tt 1.}] In the above, we have done the computation using a normal
ordered vertex. However this point is inessential. 
If we use a vertex without normal ordering, then we will
get additional terms as shown in figure 3. These diagrams arise from the 
self-contractions of the vertex and will give additional
contributions to $\cM^{(2)}$. However it is easy to see that there will not be
any noncommutative phase factor involved for these terms. 
Because of this,  they will not modify the
microcausality condition. The modification to the microcausality
condition arises solely from the nonplanar diagrams in figure 1.

\item[{\tt 2.}] Another effect of nonlocality in noncommutative 
field theory is 
the emergence of IR singularity from integrating out UV degree of
freedoms. This mixing of scales is called IR/UV mixing phenomena
\cite{iruv1,iruv2}. IR/UV mixing implies a breakdown of the standard 
Wilsonian effective description. IR/UV mixing may also results in
instability of the perturbative vacuum. This happens for example for the
case of noncommutative QED \cite{ncqed} and for the $\phi^3$ theory
\cite{iruv1} we considered. 
\footnote{The $\phi^3$ theory 
suffers also from the usual instability that it does not has 
a stable vacuum. However this is a separate issue.}
Since the vacuum is unstable, in principle our perturbative analysis
is not valid.
However our analysis can be straightforwardly 
extended to other theory where this is not a problem, 
for example $\phi^4$ in 4-dimension. 
It is clear from our analysis above that the
modification of the light cone to light wedge is due to the inclusion
of the noncommutative phase factor $P$ which appears in \eq{Hnp} for 
the nonplanar diagram. For the $\phi^4$ theory, there will
be a different $f$. However the precise form of $f$ is unimportant 
for our argument. Our analysis above go through.  
Therefore we conclude that  the modification from light cone 
to light wedge is quite  general and is valid even if 
there is  IR/UV mixing so long as the IR/UV mixing effect does not
result in instability of the vacuum.


\item[{\tt 3.}] Obviously supersymmetrizing the theory will not change our result. 
Again having more particles circulating in the loop  will only  
modify the form of the function $f$, which is not important for our
argument.

\item[{\tt 4.}] Using the equations \eq{M-gen} and \eq{G-gen}, 
our analysis can be readily generalized to higher order. 
In general a noncommutative phase factor will appear  for the nonplanar diagrams.  
For those nonplanar diagrams which has a phase
factor which intertwines the momenta in the
set $I$ in \eq{G-gen}, our analysis can be applied straightforwardly
and the light wedge is obtained. 
For example for the planar contribution in figure 2, we have the
noncommutative counterpart in figure 4 which give rise to
the light wedge due to  the phase factor  
$e^{i \vq_2 \th \vq_3}$ associated with the diagram.
 
\item[{\tt 5.}] Theory with lightlike noncommutativity  is 
unitary perturbatively \cite{lightlike} 
and presents another interesting case. 
For the case say $\th:=\th^{i-} \neq 0$, 
one quantize the theory  using the light cone 
time $x^+$ as the time coordinate. 
Since $\theta^{i+} =0$, the theory is local in
the light cone time and the operator formalism is well defined. 
Our analysis can be applied directly and we find
\begin{equation}
H_{\rm np}(q_i^+,x)\sim e^{i\frac{\th^2}{2x^+}q_2^+ q_3^+ 
(q_2^+ +q_3^+)}.
\end{equation}
This implies that the commutator vanishes only for $x_1^+ = x_2^+$. 
This agrees precisely 
with the Galilean spacetime symmetry which is in residual in 
the presence of $\th^{i-} \neq 0$.

\end{itemize}

\begin{figure}[ht] \label{higher-order-np}

\psfrag{p1}{\LARGE $p_1$}
\psfrag{q2}{\LARGE $q_2$}
\psfrag{q3}{\LARGE $q_3$}
\psfrag{q4}{\LARGE $q_4$}
\psfrag{q5}{\LARGE $q_5$}
\psfrag{qn}{\LARGE $q_n$}
\psfrag{cdots}{\LARGE $\cdots$}
\psfrag{a}{\LARGE (a)}
\psfrag{b}{\LARGE (b)}
\begin{center}
{\scalebox{0.65}{\includegraphics{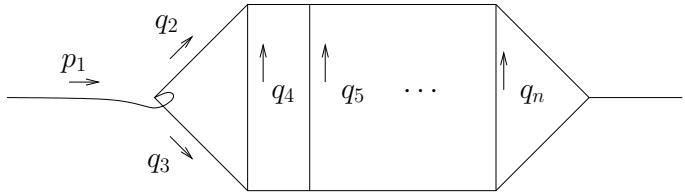}}}
\end{center}
\caption{Higher order nonplanar contributions }
\end{figure}

In conclusion, our analysis and result is applicable to general
noncommutative theories, so long as the theory admits nonplanar diagrams
and the perturbative vacuum is not instabilized by the IR/UV mixing effect.


\section{Energy condition in noncommutative geometry}

Another way to study the causality issue is to consider energy condition.
For review, see \cite{hawking}, also \cite{wald}. 
For applications in string theory, see for example \cite{gibbons}. 
Let us first consider the dominant energy
condition in details. The discussion for the other energy conditions is similar.
An energy-momentum tensor $T_{\mu\nu}$ is said to satisfy the
dominant energy condition if for every future directed time like vector
$t^\mu$, $T_{\mu\nu} t^\n$ is future directed, time like or null.
One can show that when it is satisfied,  the velocity of the energy-flow 
cannot exceed the speed of light. 

To warm up, let us first consider an example in commutative theory
\begin{equation}
\mathcal{L}=\frac{1}{2}(\partial \phi)^2-V(\phi)
\end{equation}
with $V(\phi)=\frac{1}{2}m^2\phi^2+V_{ \rm int}$.
The energy-momentum tensor is given by
\begin{equation} \label{Tmn}
T_{\mu\nu}=\partial_\mu\phi\partial_\nu\phi-g_{\mu\nu}
\mathcal{L} .
\end{equation}
We now show that the dominant energy condition is satisfied if $V>0$.
Let $t^\mu$ be timelike: $t^\mu t_\mu >0$ and future directed: $t^0 >0$.
Then
\begin{equation}
u_\mu :=  T_{\mu\nu}t^\nu =\partial_\mu\phi(\partial\phi\cdot t)-t_\mu \mathcal{L}
\end{equation}
and
\begin{equation} \label{uu1}
u_\mu u^\mu =(\partial\phi)^2 V(\phi)+t^2\mathcal{L}^2  .
\end{equation} 
Thus $u_\mu u^\mu \geq 0$ if $V(\phi) >0$. For example, if 
$m^2 >0$ and $V_{\rm int} >0$. As for $u_0$, we have
\begin{equation} \label{uu2}
u_0 =  \biggl[ \frac{1}{2} (\del_0 \phi)^2 + \frac{1}{2} 
(\del_i \phi)^2 +V\biggr]  t^0 + \del_0 \phi \del_i t^i .
\end{equation}
$t$ future directed and timelike implies $t^0 > |\bt|$. If $V>0$ also, then 
\begin{equation}
u_0 >  |\bt| \biggl[ \frac{1}{2} (\del_0 \phi)^2 + \frac{1}{2} (\del_i \phi)^2 \biggr] 
+ \del_0 \phi \del_i t^i > |\del_0 \phi| \biggl[ |\bt| |\del_i \phi| \pm t^i \del_i \phi 
\biggr] \geq 0
\end{equation}
is future directed.

Now consider the noncommutative $\phi^4$ theory
\begin{equation}
\mathcal{L}=\frac{1}{2}\left(\partial_\mu \phi\right)^2-\frac{1}{2}
m^2\phi^2-\frac{g}{4!}\phi*\phi*\phi*\phi . 
\label{action1} 
\end{equation}
Alternatively, we can also consider the Lagrangian which 
gives the same action
\begin{equation}
\mathcal{L}=\frac{1}{2}\partial_\mu \phi*\partial^\nu \phi
-\frac{1}{2}
m^2\phi*\phi-\frac{g}{4!}\phi*\phi*\phi*\phi.\label{action2}
\end{equation}
First, we need to construct the symmetric and conserved energy-momentum
tensor. This problem has been considered in  \cite{EM1}
and the authors found that they cannot be satisfied simultaneously.
Here we show that one can construct such an energy-momentum  tensor 
if one allow it to be path-dependent.  
This is  also the case in noncommutative Yang-Mills theory \cite{EM0}.
We define the energy momentum tensor
\begin{equation}
\hat{T}_{\mu\nu}=\partial_\mu \phi\partial_\nu \phi
-g_{\mu\nu}\left[\frac{1}{2}(\partial \phi)^2-\frac{1}{2}
m^2\phi^2-\frac{g}{3!}\int_{C} dy^\l \; 
\partial_\lambda
\phi(y) (\phi*\phi*\phi)(y) \right].\label{em1}
\end{equation}
Here the integration is carried out along an arbitrary path
$C$ connecting from a reference point $x_0$ to $x$.
Obviously, $\hat{T}_{\mu\nu}=\hat{T}_{\nu\mu}$ and also
$\hat{T}_{\mu\nu}$ is real.
Using equation of motion, it is easy to verify that
\begin{equation}
\partial^\mu \hat{T}_{\mu\nu}=0.
\end{equation}
For the alternative choice of the Lagrangian (\ref{action2}),
we can consider the path-dependent energy-momentum tensor
\begin{align}
\hat{T}_{\mu\nu}=&\frac{1}{2}\partial_\mu \phi *\partial_\nu \phi
+\frac{1}{2}\partial_\nu \phi*\partial_\mu \phi
-g_{\mu\nu}\biggl[\frac{1}{2}\partial_\lambda \phi*
\partial^\l \phi-\frac{1}{2}
m^2\phi*\phi \nn \\
&-\frac{g}{3!2}\int_{C} dy^\l \;
\Bigl( 
(\partial_\lambda \phi*\phi*\phi*\phi)(y)+
(\phi*\phi*\phi*\partial_\lambda \phi)(y)\Bigr)  
\biggr].\label{em2}
\end{align}
Again, it is easy to show that
$\hat{T}_{\mu\nu}$ is real, symmetric and conserved.

Let us now check the dominant energy condition in the noncommutative
case. Note that \eq{em1}  is 
of the same form as \eq{Tmn}. Thus we arrive at \eq{uu1} and \eq{uu2} 
again. However now the analogy of $V_{\rm int}$ is given by the 
path-dependent terms in (\ref{em1})
and (\ref{em2}) and they are not positive definite for non-zero
noncommutativity. One can also show that \eq{em2} does not satisfy the
dominant energy condition.
It is straightforward to generalize the argument for other
noncommutative field theories.  
Therefore we conclude that the dominant energy condition is generally 
violated in noncommutative theory.
It is  also easy to verify that none of the standard form of energy
condition (e.g. null, strong, weak) are satisfied. On the other hand, if
we naively integrating out the whole noncommutative subspace, then
energy conditions are satisfied. This is reminiscent of the averaged
energy condition \cite{tipler}.

Given that the causal structure of the noncommutative theory is modified, 
it is not surprising that the usual form of energy condition is
violated. The real challenge and really interesting problem is to derive 
the appropriate generalization of the energy condition 
for noncommutative spacetime. 

\section{Discussion}

In this paper, we have shown that the causality condition in noncommutative
field theory is generally modified from a light cone to a light wedge.
The phase factor $e^{iq_2\theta q_3}$ in \eq{J}, which arises due to the
nonlocal nature of the noncommutative interaction, is crucial to modify
the large $q_i^+$ behaviour of the kernel $H_{\rm np}(q_i^+,x)$ and hence
the modified causality condition. In contrast, a local Lorentz symmetry
breaking interaction will generally not modify the light cone.
Nonlocality of interaction plays the key role. 
See also \cite{zizzi1} for other considerations on how  
noncommutative geometry may modify the causality condition.

Based on the modified dispersion relation in the noncommutative theory 
at nonzero temperature, 
the authors of \cite{karl} show that the low momenta modes have 
superluminous group velocity. 
The authors of \cite{ncsoliton} constructed 
and studied the classical dynamics of solitons in noncommutative
gauge theories. They found that these solitons 
can travel with speed faster than light (with no
upper bound in the noncommutative direction). 
However as these authors argued, 
since one cannot  
send information backward in time, so this does not violate
causality. One may also deduce the light wedge in this
setting. Our result of having the light cone changed to the light wedge
is consistent with their result. However 
the analysis and result here is more general since it applies 
without the need
of specific modified dispersion relation nor
of the existence and the construction of solitons with specific properties. 

In \cite{chaichian1}, a new form of Wightman functions, which are defined
as vacuum expectation value of star-product of fields, have been
considered. The motivation was that the new form of Wightman functions 
contain the
noncommutativity explicitly. On the other hand, as the authors argued, 
on applying the reconstruction theorem the usual Wightman functions 
will lead to
commutative field theory (which are invariant under the smaller Lorentz
group), but not to noncommutative field theory since there is no trace
of the noncommutativity parameter $\th$ in the usual Wightman
function. We remark that this is not the case. The
Wightman functions can depend on $\th$ just as they depends on the
coupling constants. This can be easily demonstrated in perturbation theory, as
is clear from our calculations. 
Also the authors proposed another form of microcausality condition where
the star-commutator of fields vanishes at separation which is spacelike
with respect to the light wedge. We note that, due to momentum
conservation, it is
\begin{equation}
\bra{\a} [\phi(x_1),\phi(x_2)]\ket{\b} = 
\bra{\a} [\phi(x_1),\phi(x_2)]_*\ket{\b},
\end{equation}
for states $\ket{\a}, \ket{\b}$ with the same momentum. Thus as far as
matrix elements of the above type are of concern, there is no difference
in considering the standard commutator or the
star-commutator. The usual commutator is smart enough to know about
the modified causality condition in noncommutative geometry.

As we explained, the emergence of the modified causality condition is
characteristic of noncommutative geometry and is generic and independent
of the details of the type of noncommutative interaction. 
One can expect that this drastic change of the notion
of causality to be a rather clear phenomenological signal for
noncommutative geometry. See for example \cite{harvey,pheo, susy}
for some discussions of the phenomenological aspect of noncommutative 
field theory. 
If the universe was noncommutative at the early stage, 
region that one would traditionally taken to be not in causal contact
(outside the light cone) may indeed be causally related according 
to the light wedge. This may offer an alternative scenario to inflation 
for explaining the horizon problem of the universe. 
For other applications of noncommutative geometry to cosmology, 
see for example \cite{inflat}.
 
In this paper, we have seen that nonlocality of noncommutative
geometry leads to a modification of the light cone. String theory is
another nonlocal theory. The effect of interaction on the string
light cone \cite{martinec} has been studied in \cite{flat} for the case
of flat string theory and \cite{pp} for the pp-wave string theory. The
result is that the light cone is modified in the flat case, but not in
the pp-wave case \cite{pp}. This is related to the fact that the pp-wave
3-string interaction is more localized compared to the flat space
3-string interaction. It is important to identify and understand better the
nonlocal effects in string theory. This should help us to understand
better the nature of the theory of quantum gravity, which is  
believed to be nonlocal also. 

The modification of the light cone in  noncommutative field theory is
intriguing. As far as causality is of concern, no trace of the
noncommutative coordinates enter. Our result suggests that a
reformulation of the noncommutative theory in terms of one living in lower
dimensional spacetime maybe able to capture the 
causal aspects of the physics more accurately. 
This possibility of formulation in terms of lower dimensional degree
of freedom reminds us of the holographic 
property of quantum gravity \cite{hol}. 
The dynamical reduction of spacetime dimensions at short distance scale 
is also seen in the very interesting recent works \cite{ajl}.
Since quantum gravity effects is believed to quantize the spacetime and
make it noncommutative, noncommutativity geometry may   
play a key role in explaining holography. 
See also \cite{ho-li,zizzi2} for more considerations 
in support of this view.


\section*{Acknowledgements} 
CSC acknowledges the support of EPSRC through an advanced fellowship.
TI acknowledges supports from the grants from JSPS (Kiban C and Kiban B). 
The authors wish to thank Koryu Kyokai for the grant of Japan-Taiwan
science collaboration, which helped us meet and collaborate.

\appendix

\section{Light cone for free theory}
Consider real scalar field in $(d+1)$-dimensions. It is easy to
calculate that
\begin{equation}
[\phi(x), \phi(0) ] = D(x) -D(-x), \quad D(x) :=
\int \frac{d^d \bq}{(2\pi)^d}\frac{1}{2 \o_q} e^{-i q\cdot x} 
\end{equation}
Using  \eq{measure},
it is easy to rewrite $D$ as
\begin{equation}
D(x)=  \int_0^\infty  dq^+ G(q^+,x),\quad
G(q^+,x) :=\frac{1}{4 \pi  q^+}
\int \frac{ d^{d-1}\vq}{(2\pi)^{d-1}}  e^{- i q\cdot x} .
\end{equation}
The kernel $G(q^+,x)$ can be easily evaluated and we obtain
\begin{equation} \label{com-tree}
[\phi(x), \phi(0) ] =\int_0^\infty  dq^+ [G(q^+,x) - G(q^+,-x)],
\end{equation}
\begin{equation}
G(q^+,x) = \frac{1}{4 \pi  q^+} (\frac{i q^+}{2 \pi x^+})^{\frac{d-1}{2}} 
\exp \bigg[ -i q^+ \frac{\d x^2}{2x^+} 
- \frac{i m^2 x^+}{2 q^+} \bigg].
\end{equation}
The convergence properties of the $q^+$ integral depends crucially on
the sign of $\d x^2 =2 x^+x^- - \vx^2$.
Without loss of generality, we assume $x^+ >0$. 
If $\d x^2 <0$, then we can rotate the contours in
\eq{com-tree} as follows without getting any extra contribution:
\bea
&& q^+ \to i q^+ \quad \mbox{in the first integral}, \nn\\
&& q^+ \to -i q^+ \quad \mbox{in the second integral}.
\eea
Moreover since
\begin{equation}
G(iq^+,x) \sim e^{q^+ \frac{\d x^2}{2x^+} -  \frac{m^2 x^+}{2 q^+}}
\end{equation}  
the integral converges for $q^+ \to 0$ and
$q^+ \to \infty$   if $\d x^2<0$. 
Hence the two terms in \eq{com-tree} cancels and the
commutation vanishes if $\d x^2<0$. On the other hand, if $\d x^2 >0$, we may perform
the opposite contour rotation. The integral converges for $q^+ \to
\infty$ but not for  $q^+ \to 0$. 
Therefore we obtain that the commutator vanishes iff   $\d x^2 <0$.



\end{document}